\documentclass[aps,pra,preprint,superscriptaddress,showpacs]{revtex4}

\usepackage{amsmath,amssymb,enumerate}
\usepackage[latin1]{inputenc}
\usepackage{boxedminipage}

\newcommand{\qu}[1]{\ensuremath{|#1\rangle}}

\begin{document}


\title{Quantum authentication scheme based on algebraic coding}


\author{Rex A. C. Medeiros}
\email[e-mail: ]{rex@dee.ufcg.edu.br}
\author{Francisco M. de Assis}
\email[e-mail: ]{fmarcos@dee.ufcg.edu.br}
\author{Bernardo L. J\'unior}
\email[e-mail: ]{lula@dsc.ufcg.edu.br}
\author{A\'ercio F. Lima}
\email[e-mail: ]{aerlima@df.ufcg.edu.br}
\affiliation{Universidade Federal de Campina Grande, Av. Aprígio Veloso, 882, 58.109-970 Campina Grande, Brazil}


\date{\today}

\begin{abstract}
This paper presents a simple, but efficient class of non-interactive protocols for quantum authentication of $m$-length classical messages. The message is encoded using a classical linear algebraic code $C[n,m,t]$. 
We assume that Alice and Bob share a classical secret key $x_{AB}$, of $n$ bits. Alice creates $n$ qubits based on the codeword and the key, that indicates the bases used to create each qubit. The quantum states are sent to Bob through a noiseless quantum channel. We calculate the failure probability of the protocol considering several types of attacks.
\end{abstract}


\maketitle

\section{Introduction}
Authentication is a procedure to verify that a received message comes from a certain entity, and have not been altered. Classical cryptography describes several techniques to implement authentication. The Message Authentication Code (MAC) presumes the existence of a secret key shared among the two parts, A (Alice) and B (Bob). The coding algorithm generates a tag, known as a cryptographic checksum, which is a function of the message and the key. The tag is attached to the message. The recipient performs the same calculation on the received message, using the same secret key, to generate a new tag to be compared to the received tag. Identical tags indicate that the received message is authentic~\cite{WSta:98}.

The discovery and formalization of quantum mechanics during the last century motivated studies in the fields of computation and information theories~\cite{NC:2000,Spi:96,BS:98}. Effects like entanglement and the discovery of EPR pairs made possible quantum states teleportation~\cite{BBCJ:93}. Some problems computationally intractable in the classical world, as factorization, are solved using polynomial algorithms running on a quantum computer. The development of such technology would make unfeasible, for example, public key cryptographic systems, whose security is based on the inefficiency of classical factorization algorithms~\cite{WSta:98}. One of the most interesting applications of quantum information theory is quantum cryptography. In 1970, Wiesner showed that quantum mechanics properties could be used for such end, but his work was only published in 1983~\cite{Wies:83}. Later, Bennett and Brassard described a quantum key distribution protocol known as BB84~\cite{BB84:84}. There exist several proofs of unconditional security of BB84~\cite{Maye:98,HKLo:99,ShPr:00}, even against any collectives attacks~\cite{BBBG:98}.

Until the last decade, the expression ``quantum cryptography'' referred basically to protocols for quantum key distribution (QKD). Recently, several researches have been made in the sense of applying quantum mechanics resources in the resolution of others problems related to the data security. The first works deals to the key verification~\cite{ZeZh:00} and user authentication~\cite{DuHa:99,ZeGu:00,JeSc:00}. Key verification consists of assuring the legitimacy of the two parts involved in a key distribution scheme, and that the established key is authentic. User authentication, also called user identification, allows a system to determine the users identity that wants to use it.

Curty and Santos~\cite{CuSa:01} proposed a protocol to quantum authentication of unitary-length classical messages (bit). As for the secret key, they use a maximally entangled EPR pair previously shared between Alice and Bob. For each message, an EPR pair is used. Alice needs to generate, through a unitary operation, a quantum state, called quantum tag, which depends on the qubit that represents the classical bit and her part of the EPR pair. For the types of attacks discussed, the probability $P_d$ that Eve deceives Bob was $0.5 \leq P_d <1$, depending on the choice of the unitary operation. Later, the same authors proposed a protocol to quantum authentication of unitary-length quantum messages (qubit)~\cite{CuSP:02}. The second protocol is a generalization of the first, where the quantum tag now belongs to a state space of dimension equal to or greater than the dimension of the message state space.

Recently, Barnum {\it et al.}~ \cite{BCGS:02} described a protocol to authenticate quantum messages of length $m$. They propose a scheme that both enables Alice to encrypt and authenticate (with unconditional security) an $m$ qubit message by using a stabilizer code to encode the message into $m+s$ qubits, where the failure probability decreases exponentially in the security parameter $s$. Such scheme requires a private key of size $2m + O(s)$ to be shared between Alice and Bob. To archive this, the authors proposed a protocol for testing the purity of shared EPR pairs. This protocol needs quantum circuits for the coding and decoding operations. 

In this paper we address the problem of authenticating classical messages of length $m$ transmitted over a noiseless quantum channel. We propose a non-interactive scheme that just requires preparation of quantum states into orthornormal bases, transmission and measurements of these states in the same bases. To reach a wished security level, the message should be coded using a classical linear algebraic code $C[n,m,t]$~\cite{Wick:95}. After the coding operation, Alice creates $n$ qubits based on the chosen codeword and on a secret key of $n$ bits, $x_{AB}$, previously shared with Bob. The key indicates the bases used by Alice and Bob for the creation and measurement of the qubits. Bob assumes that no forgery has taken place and that the message is authentic if the result of the measurement is a codeword $c\in C[n,m,t]$. We consider two types of attacks, the no-message and intercept-resend attacks. We calculate the probability of a eavesdropping successfully forger a message to deceive Bob. As we will show, this probability depends on the parameters $n$ and $t$ of the code $C[n,m,t]$.

\section{\label{sec:esquema}A Protocol to Quantum Authentication of $m$-Length Classical Messages}

Suppose Alice wants to send Bob a $m$-bits certified message, $k_i$, chosen from a set $K = \{k_i \} = \{0,1\}^m$. Bob, when receiving the message, should be able to infer about its authenticity, i.e., if the message was sent by Alice or not. The protocol described in this section makes use of a classical linear algebraic code $C[n,m,t]$ with parity check matrix $H$, and a noiseless quantum channel for transmission of coded messages. For each message $k_i \in K$ we associate a codeword $c_i \in C$. Participants must share a classical secret key of $n$ bits, $x_{AB}$, chosen in a random and independent way.

The authentication procedure is described as follows. Initially, Alice and Bob define two orthonormal bases for the 2-dimension Hilbert space, ${\cal Z} = \{\qu{0},\qu{1} \}$ and ${\cal X} =  \{\qu{+} = \frac{1}{\sqrt{2}} (\qu{0}+\qu{1}), \qu{-} = \frac{1}{\sqrt{2}} (\qu{0}-\qu{1}) \}$. When Alice needs to send the message $k_A $, she computes the corresponding codeword $c_A$. For each bit of $c_A$, Alice prepares a quantum state $\qu{\psi_j}$ based on the corresponding key bit. Then, if the $j$-th bit of $x_{AB}$ is 0, Alice prepares \qu{\psi_j} using ${\cal Z}$ basis, such that

\begin{equation}
\qu{\psi_j} =
\begin{cases}
    \qu{0} & \quad \text{ if the }j\text{-th bit of } c_A \text{ is } 0\\
    \qu{1} & \quad  \text{ if the }j\text{-th bit of } c_A \text{ is } 1.
\end{cases}
\end{equation}

Similarly, if the $j$-th bit of $x_{AB}$ is 1, Alice prepares \qu{\psi_j} using ${\cal X}$ basis, such that
\begin{equation}
\qu{\psi_j} =
\begin{cases}
    \qu{+} & \quad \text{ if the }j\text{-th bit of } c_A \text{ is } 0 \\
    \qu{-} & \quad \text{ if the }j\text{-th bit of } c_A \text{ is } 1.
\end{cases}
\end{equation}

After the qubits generation, Alice sends the state $\qu{\psi_j}^{\otimes n}$ to Bob through the quantum channel.

At the reception, Bob makes measurements to obtain a sequence $m_B$ of $n$ bits. For the $j$-th received qubit, Bob measures it using the basis ${\cal Z}$ or ${\cal X}$ depending on the $j-$th bit of $x_{AB}$ is 0 or 1, respectively. Because the quantum channel is perfect, Bob recognizes that the message is authentic if $m_B$ is a codeword, i.e., $m_{B}H^{T} = 0 $. Then, Bob decodes $m_B$ to obtain the authentic message. Otherwise, Bob assumes that Eve tried to send him an unauthentic message. He then discards the received message. After each transmission, Alice and Bob discard the key $x_{AB}$.

\section{\label{sec:seguranca}Security Analysis}

In this section we analyze the security of the proposed protocol, for the case of a noiseless quantum channel connecting Alice and Bob. Two types of attacks will be considered: the no-message attack and the intercept-resend attack. In the first one, Eve prepares a quantum state and sends it to Bob. In the second, Eve intercepts the qubits and performs measurements in attempting to obtain some information about the message sent by Alice. Then, Eve exploits the information gained to prepare possibly another message that she sends to Bob.

For the analysis, we consider that the linear code $C$ and the mapping $k_i \rightarrow c_i$ are publicly known, what is a realist assumption.

\subsection{No-Message Attack}

We analyze here the case where Eve precedes Alice and sends Bob a quantum state $\qu{\psi_\epsilon} = \qu{\psi_{\epsilon_j}}^{\otimes n}$ trying to impersonate Alice. Let Eve choose a message $k_E$, with associated codeword $c_E$. Because Eve does not know anything about the key $x_{AB}$, she chooses a random sequence $x_E$ to indicate the bases used to create the qubits.

To calculate the protocols failure probability $P_{f}$, we define, for each message bit, three events: $\varepsilon_1 =$ Eve chooses the same basis than Bob; $\varepsilon_2 =$ Eve chooses a different basis from Bob; and $\varepsilon_3 =$ Bob obtain, after measurement, the same bit that Eve sent. The probability $P_f$ that Eve cheats Bob is therefore,
\begin{equation}
\label{eq:pf0}
    P_{f} = (P(\varepsilon_3|\varepsilon_1)P(\varepsilon_1) + P(\varepsilon_3|\varepsilon_2)P(\varepsilon_2))^n.
\end{equation}

The first conditioned probability, $P(\varepsilon_3|\varepsilon_1)$, is equal to 1 due to the fact that, when Eve chooses the same basis that Bob uses to measure the qubit, Bob will always obtain by the measurement the same bit that Eve wished send to him. When Eve misses the basis, Bob measures the same bit sent by Eve with probability $P(\varepsilon_3|\varepsilon_2) = 1/2$. Then,

\begin{eqnarray}
\label{eq:pf}
    P_{f} &=& (1 \times \frac{1}{2} + \frac{1}{2} \times \frac{1}{2})^n \notag\\
        &=& (3/4)^n.
\end{eqnarray}

\subsection{\label{sec:ataquesubstituicao}Intercept-Resend Attack}

In this type of attack, Eve segments the quantum channel between Alice and Bob, intercepts and measures the quantum states that are being transmitted to Bob. Based on the gained information, Eve prepares another message of her interest and sends it to Bob. 

Since Eve has no information about the secret key $x_{AB}$, she must initially choose her bases sequence, called here $x_E$, to measure the qubits. We investigate here an eavesdropper strategy where Eve attempts to decode correctly the $n$ bit string resulting from measurements, to obtain the codeword sent by Alice. If Eve makes this successfully, she can exploit the gained information to partially correct her key $x_E$, forge an authentic message and send it to Bob.

Suppose Alice generates the quantum state $\qu{\psi_j}^{\otimes n}$ based on the codeword $c_A$, corresponding to the message $k_A$, and the classical secret key $x_{AB}$ shared with Bob. Initially, we calculate the probability $P_{\text{dec}}$ of Eve decodes successfully the bit string resulting from measurements to obtain $c_A$. To perform this, we employ the error correction properties of the code $C$ together with our strategy to create the quantum states. Let $m_E$ the $n$-bits sequence resulting of Eves measurements using her key $x_E$ randomly generated. Define the random variable $X =$ number of bits of Eves key $x_E $, that coincides with the bits of Alice and Bobs key $x_{AB}$. If $e = m_E + c_A$ is the error vector when Eve measures $\qu{\psi_j}^{\otimes n} $, then
\begin{eqnarray}
\label{eq:pde}
    P_{\text{dec}} &=& P(w(e) \leq t) \notag\\
        &=& \sum_{i=0}^{n-t-1} P(X=i) P(w(e) \leq t | X = i) \notag \\
    & &\quad + \sum_{i=n-t}^{n} P(X=i),
\end{eqnarray}
where $w(e)$ stands for error vector Hamming weight~\cite{Wick:95}. Because if Eves bases sequence $x_E$ matches $x_{AB}$ in $n- t-1$ or more positions, then the error vector weight is always less than or equal to $t$, i.e., $P(w(e) \leq t | X = i) = 1$ since $i \ge n-t$.

It is straightforward to see that $X$ has a binomial distribution with $p=1/2$,
\begin{eqnarray}
\label{eq:pbina}
    P(X = i) &=& \binom{n}{i} (1/2)^i (1/2)^{n-i}  \notag\\
    &=& \binom{n}{i} 2^{-n}.
\end{eqnarray}

Thus, for any realization of $X$, $X = i$, Eve knows with probability one that $i$ bits of $m_E$ are correct bits. Among $n - i $ remaining bits of $m_E$, Eve should measure correctly at least $n-i-t$ bits to be able to correct the word. But, if Eve misses the basis, she has probability $p=1/2$ of still obtain the correct bit. So, for $0 \leq i \leq n - t-1$
\begin{eqnarray}
\label{eq:pw}
    P (w(e) \leq t | X = i) &=& \sum_{h=0}^{t} \binom{n-i}{n-i-h}(1/2)^{n-i-h}(1/2)^{h} \notag\\
    &=& \sum_{h=0}^{t} \binom{n-i}{n-i-h} 2^{-(n-i)}. 
\end{eqnarray}

The probability $P_{\text{dec}}$ is then
\begin{eqnarray}
\label{eq:pdec}
    P_{\text{dec}} &=& \sum_{i=0}^{n-t-1} \sum_{h=0}^{t} \binom{n}{i} \binom{n-i}{n-i-h}2^{-(2n-i)}  \notag\\
    & & \quad + \sum_{i=n-t}^{n} \binom{n}{i} 2^{-n}. 
\end{eqnarray}

The decoding probability found above depends on the parameters $n$ and $t$ of the chosen code. Once Eve decodes successfully the codeword sent by Alice, there is an increase on the failure probability of the system. This is because Eve gains information from Alice and Bob's secret key, allowing Eve partially corrects its key $x_E$. To achieve this, she compares the bits of $m_E$ with the bits of $c_A$. Eve concludes that chose wrong bases in the positions where $m_E$ differ with $c_A$. She then flips the incorrect bits of $x_E$ to obtain a new key $x_E'$.

For the scenario described above, it is possible to calculate the systems failure probability. Assume that when Eve decodes a message correctly, she corrects $t$ positions of $x_E$. Define the events: $\varepsilon_i =$ Eve guesses $i$ positions of $x_{AB}$ and decodes $m_E$ correctly; $\varepsilon =$ Bob accepts the received message as authentic. If Eve chooses a message to transmit and creates the quantum state $\qu{\psi_E}^{\otimes n}$ based on the corresponding codeword and the corrected key $x_E$, the protocols failure probability $P_f'$ can be written as
\begin{equation}
\label{eq:pff}
    P_f' = \sum_{i=0}^{n-t-1} P(\varepsilon |\varepsilon_i) P(\varepsilon_i) + \sum_{i=n-t}^{n} P(\varepsilon_i).
\end{equation}

When Eve guesses more than $n-t-1$ bases, she decodes the message correctly, and she can correct entirely its key to obtain $x_E' = x_{AB}$. Therefore, Eve always deceives Bob, i.e., $P(\varepsilon|\varepsilon_i) = 1$ for $n-t \leq i \leq n$.

To calculate the conditioned probability $P(\varepsilon | \varepsilon_i)$, it is enough to notice that $i$ bits of  $x_{AB}$ were initially correct and $t$ bits were corrected. Therefore, there exists $w(x_{AB}+x_E) = n-t-i$ incorrect bits in $x_E$. Although Eve misses the basis, there is a probability equals to $1/2$ of Bob measures the same bit sent by Eve, so that

\begin{equation}
P(\varepsilon |\varepsilon_i) = 2^{-(n-t-i)}.
\end{equation}

The probability $P(\varepsilon_i)$ it was previously discussed [Eqs. (\ref{eq:pw}) and (\ref{eq:pdec})]. For $0 \leq i \leq n - t-1$,
\begin{equation}
    P(\varepsilon_i) =  \sum_{h=0}^{t} \binom{n}{i} \binom{n-i}{n-i-h}2^{-(2n-i)},
\end{equation}
and for $i \ge n - t$,
\begin{equation}
    P(\varepsilon_i) =   \binom{n}{i} 2^{-n},
\end{equation}
so that the probability $P_f'$ is
\begin{eqnarray}
\label{eq:pflinha}
    P_{f}' &=& \sum_{i=0}^{n-t-1} \sum_{h=0}^{t} \binom{n}{i} \binom{n-i}{n-i-h}2^{-(3n-2i -t)}  \notag  \\
    & &\quad + \sum_{i=n-t}^{n} \binom{n}{i} 2^{-n}. 
\end{eqnarray}

\section{\label{sec:resumo}Protocol Summary}

Considering that Alice and Bob share a random secret key $x_{AB}$ and they agree on a linear algebraic code $C[n,m,t]$, the proposed protocol for quantum authentication of classical messages can be summarized as follows:

\noindent
\begin{boxedminipage}[t]{\linewidth}
\begin{description}
\item[\textbf{1.}] Alice chooses $c_A \in C$ corresponding to $k_A$.
\item[\textbf{2.}] Alice creates $n$ qubits in the bases $\cal Z$ or $\cal X$, depending on $x_{AB}$. She sends the qubits through the quantum channel.
\item[\textbf{3.}] Bob chooses the bases used in the measurements according with $x_{AB}$. The measurement results is a $n$-bits sequence $m_B$.
\item[\textbf{4.}] Bob performs a parity test on $m_B$. Case $m_BH^T \neq 0$, the message is discarded (Eve interfered in the channel). If $m_B$ passes the parity test ($m_BH^T = 0$), Bob obtains the message $k_A$ decoding $m_B = c_A$.
\end{description}
\end{boxedminipage}

\section{Discussion}

According with the analyses presented in the section~\ref{sec:seguranca}, the security of our protocol depends on the parameters $n$ and $t$ of the linear algebraic code $C[n,m,t]$ chosen. Moreover, the failure probabilities can be made as small as wished. To have an idea of such security, we calculated the probabilities discussed for several binary BCH codes of lengths $n=63$ and $n=127$~\cite{Wick:95} (Table~\ref{tab:bch}).

\begin{table}
\caption{\label{tab:bch}Security of the protocol for some binary BCH codes.}
\begin{ruledtabular}
\begin{tabular}{cccc}
$\mathbf{C[n,m,t]}$  & $\mathbf{P_{f}}$ & $\mathbf{P_{Dec}}$ & $\mathbf{P_f'}$ \\
\hline $\mathbf{C[63,57,1]}$ & $1.3\times 10^{-8}$& $4.1\times 10^{-15}$ & $2.8\times 10^{-13}$\\

\hline $\mathbf{C[63,51,2]}$ & $1.3\times 10^{-8}$& $4.4\times 10^{-16}$ & $5.5\times 10^{-13}$\\

\hline $\mathbf{C[63,18,10]}$ & $1.3\times 10^{-8}$& $3.1\times 10^{-9}$ & $3.2\times 10^{-9}$\\

\hline $\mathbf{C[63,10,13]}$ & $1.3\times 10^{-8}$& $3.7\times 10^{-7}$ & $3.7\times 10^{-7}$\\

\hline $\mathbf{C[127,120,1]}$ & $1.4\times 10^{-16}$& $3.0\times 10^{-32}$ & $2.4\times 10^{-26}$\\

\hline $\mathbf{C[127,113,2]}$ & $1.4\times 10^{-16}$& $5.0\times 10^{-34}$ & $4.8\times 10^{-26}$\\

\hline $\mathbf{C[127,36,15]}$ & $1.4\times 10^{-16}$& $1.0\times 10^{-20}$ & $1.1\times 10^{-20}$\\

\hline $\mathbf{C[127,22,23]}$ & $1.4\times 10^{-16}$& $1.8\times 10^{-14}$ & $1.8\times 10^{-14}$\\
\end{tabular}
\end{ruledtabular}
\end{table}

When the message to be send is a random sequence of bits, the classical message authentication codes (MAC) presents only a computational security, even when a larger key is used to produce the authentication block~\cite{WSta:98}. The class of protocols described here presents an information theoretic security, rather than based on computational assumptions.

The size of the used secret key is another important aspect. In general, if $R = m/n$ is the rate of the linear code $C[n,m,t]$, the length of the key will be $1/R$ times the length of the message. For example, for the code  $C[127,120,1]$, it is only necessary a key whose length is $(127/120) \cong 1,06$ times the length of the message to guarantee, in the worst case, a failure probability of $1.4\times 10^{-16}$.

Moreover, there exists a possibility of reusing the secret key for Alice and Bob, since in quantum systems it is possible to identify an attempt to perturbing the states transmitted through the channel. If the quantum channel can be considered perfect and Bob receives an authentic message, he can conclude that no eavesdropper was present. Then, the secret key can be reused without compromise the security of the protocol.

Comparatively to others quantum schemes to authenticate classical messages present in the literature, the protocol described here has advantages in terms of simplicity and use of quantum resources. The Curty and Santos's protocol needs a quantum operation to generate a quantum tag to be send attached to the message. Moreover, the security of such protocol depends on the choice of the unitary operation. However, the authors did not show the existence of a optimum unitary operation that minimize the failure probability~\cite{CuSa:01}.

A disadvantage of our scheme is the use of a classical secret key. This means that it is possible to read or copy the key by a third part during storage process, without being detected. This problem will only be solved with the improvement of equipments to storage quantum states.

\section {Conclusions}
In this work we presented a simple, but efficient non-interactive scheme for quantum authentication of $m$-length classical messages. The described protocol make uses of a linear algebraic code $C[n,m,t]$ to encode the message and a classical secret key of $n$ bits. The quantum states are created based on the codeword, where the key bits are used to choose the bases. Then, the qubits are transmitted through a quantum channel. 

According with quantum mechanics theory and considering the systematic adopted in the quantum states creation, the protocols failure probabilities were calculated. In general, these probabilities depend on the choice of the parameters $n$ and $t$ of the code, and we can make them as small as desired. The security of the proposed protocol does not depend, therefore, on computational assumptions.


%



\begin{acknowledgments}
The authors trank the Brazilian National Council for Scientific and Technological Development (CNPq) for support (CT-INFO Quanta, under grants 552254/02-9).
\end{acknowledgments}


\end{document}